\documentstyle[prl,aps]{revtex}
\begin{document}
\title
{Comment on ``Velocity at the Schwarzschild Horizon Revisited'' by I. Tereno}

\author{Abhas Mitra}
\address{Theoretical Physics Division, Bhabha Atomic Research Center,\\
Mumbai-400085, India\\ E-mail: amitra@apsara.barc.ernet.in}


\maketitle
\begin{abstract}

\end{abstract}
\vskip 1cm

We have recently shown that even when we describe a Sch. BH by the Kruskal
coordinates, the radial geodesic of a material particle, which must be
timelike for any finite $r$ becomes null at the Event Horizon
($r=2M$)\cite{1}.

To see this one has to recall the structure of the Kruskal transformations:(Sector I):

\begin{equation}
u=f_1(r) \cosh {t\over 4M}; \qquad v=f_1(r) \sinh {t\over 4M};
\qquad r\ge 2M
\end{equation}
where
\begin{equation}
f_1(r) = \left({r\over 2M} -1\right)^{1/2} e^{r/4M}
\end{equation}
It would be profitable to note that
\begin{equation}
{df_1\over dr} = {r\over 8M^2} \left({r\over 2M} -1\right)^{-1/2} e^{r/4M}
\end{equation}
And for the region interior to the horizon (Sector II), we have
\begin{equation}
u=f_2(r) \sinh {t\over 4M};\qquad v=f_2(r) \cosh {t\over 4M};
\qquad r\le 2M
\end{equation}
where
\begin{equation}
f_2(r) = \left(1- {r\over 2M}\right)^{1/2} e^{r/4M}
\end{equation}
and
\begin{equation}
{df_2\over dr} = {-r\over 8M^2} \left(1- {r\over 2M}\right)^{-1/2} e^{r/4M}
\end{equation}
Given our adopted signature of spacetime ($-2$), in terms of $u$ and $v$, the metric for the entire spacetime is
\begin{equation}
ds^2 = {32 M^3\over  r} e^{-r/2M} (dv^2 -d u^2) - r^2 (d\theta^2
+d\phi^2 \sin^2 \theta)
\end{equation}
The metric coefficients are regular everywhere except at the intrinsic
singularity $r=0$, as is expected. Note that, the angular part of the
metric remains unchanged by such transformations and $r(u,v)$ continues to
signal its intrinsic spacelike nature. Though, in either region we have
\begin{equation}
u^2-v^2= \left({r\over 2M} -1\right) e^{r/2M}
\end{equation}

the functional form of $u/v$ is in general different in the two regions:

\begin{equation} {u\over v} = \coth {t\over 4M}; \qquad r
\ge 2M \qquad Sector I
\end{equation}
and
\begin{equation}
{u\over v} = \sinh {t\over 4M}; \qquad r \le 2M \qquad Sector II
\end{equation}
and

so that
\begin{equation}
u^2-v^2 > 1; \qquad  u/v >\pm 1; \qquad r >2M,
\end{equation}
\begin{equation}
u^2-v^2 \rightarrow 0; \qquad u =\pm v; \qquad r= 2M
\end{equation}
and
\begin{equation}
u^2-v^2 <0; \qquad u/v < \pm 1; \qquad r <2M
\end{equation}
 First we focus attention on the region $r\ge
2M$ and differentiate Eq.(1) to see
\begin{equation}
{du\over dr} = {\partial u\over \partial r} + {\partial u\over \partial t}
{dt\over dr}= {df\over dr} \cosh {t\over 4M} + {f\over 4M} \sinh
{t\over 4M}{dt\over dr}
\end{equation}
Now by using Eq. (4-6) in the above equation, we find that
\begin{equation}
{du\over dr} = {ru\over 8M^2} (r/2M -1)^{-1} + {v\over 4M} {dt\over dr}; \qquad
r\ge 2M
\end{equation}
and
\begin{equation}
{dv\over dr} = {rv\over 8M^2} (r/2M -1)^{-1} + {u\over 4M} {dt\over dr};
\qquad r\ge 2M
\end{equation}
By dividing equation (15) by (16), we obtain
\begin{equation}
{du\over dv} = {{ru\over 2M}  + v {dt\over dr} (r/2M-1) \over
{rv\over 2M}  + u {dt\over dr} (r/2M -1)}
\end{equation}
Similarly, starting from Eq. (5), we end up obtaining a form of
$du/dv$ for the region $r <2M$ which is exactly similar to the
foregoing equation. Now, by using Eq.(12) ($u=\pm v$) in
 Eq. (17), we promptly find that
\begin{equation}
{du\over dv}\rightarrow {{\pm r\over 2M}  +  {dt\over dr} (r/2M-1) \over
{ r\over 2M}  \pm {dt\over dr} (r/2M -1)} \rightarrow \pm 1;\qquad r\rightarrow 2M
\end{equation}
Thus, we are able to find the precise value of $du/dv$  at the EH in a
most general manner {\em irrespective of the precise relationship} between
$t$ and $r$.

Then it follows from {\em radial part} of the Kruskal metric ($d\theta
=d\phi =0$) as
\begin{equation}
ds^2 = {32 M^3\over  r} e^{-r/2M} dv^2 \left[1- \left({du\over
dv}\right)^2\right] =0 ; \qquad r=2M
\end{equation}

Our work has now been reinterpreted by Tereno\cite{2} in terms of
the physical 3-velocity measured by a Kruskal observer (at $r
=2M$).
He reexpresses our Eq. (17) as:

\begin{equation}
V = {du\over dv} = {u + v {dt\over dr} (r/2M-1) \over v + u
{dt\over dr} (r/2M -1)}
\end{equation}

And since it is clear from Eq. (12) that $u/v \rightarrow \pm 1$ as $r
\rightarrow 2M$, $V \rightarrow \pm 1$.  But Tereno thinks otherwise.
After a somewhat convoluted exercise,  he arrives at an
expression

\begin{equation}
V = - {\epsilon - \delta/E^2\over \epsilon + \tanh (t/4M) \delta/E^2}
\end{equation}
where $E^2$ is finite and

\begin{equation}
\epsilon =1 -\tanh (t/4M), \qquad \delta = (r-2M)/2r
\end{equation}
Now he concludes that the value of $\mid V\mid <1$! We assert
that this conclusion arises from {\em his inability to work out
the limiting value of fractions}. For instance, by using the fact
the Sch. time $t
\rightarrow \infty$ as the EH is approached
one can find
\begin{equation}
\tanh {t\over 4M} = \coth {t\over 4M} \rightarrow 1; \qquad r\rightarrow 2M
\end{equation}

Then since
\begin{equation}
\delta \rightarrow 0; \qquad r \rightarrow 2M
\end{equation}
  Eq. (21) reduces to

 \begin{equation}
 V \rightarrow - {\epsilon\over \epsilon} \rightarrow -1 \qquad
 r\rightarrow 2M
 \end{equation}
 even though $\epsilon \rightarrow 0$ in this limit.
 Had we evaluated the limit on $\delta \rightarrow 0$ first, on the other
hand, we would have obtained $V \rightarrow +1$, so that eventually, $V
\rightarrow \pm 1$ in agreement with Eq.(18). And the fact that $V$ can be
both +ve as well as -ve is no virtue for the underlying theory. In fact,
even the innocuous Sch. spacetime has got its quota of anomalous behaviour
(Sch. singularity) if it is assumed that $M >0$; in particular, it can be
shown that

\begin{equation}
{dr\over dt} > 0; \qquad for~ r < 2M
\end{equation}

To see this explicitly just differentiate the $r-t$ relationship (pp. 824
of Gravitation by Misner, Thorne and Wheeler) or find from Eq. (11) of Tereno.
All such anomalies are linked with the idea of EH and BHs, and they vanish
when we realize that, mathematically, $M=0$. And Physically, at a finite
proper time, there is no EH or BH but only Eternally Collapsing
Configurations (ECC).

 It is trivial to
verify that his conclusion that the velocity addition law of Sp. Theory of
Relativity breaks down when the individual velocities approach unity
{\em arises from his inability to work out limiting values of fractions}.
 He should execute
one limit at a time and not merely put the $V_1 =V_1 =1$ at one go. In
this way he would arrive at a relative velocity $ V =\pm 1$, where the two
signs correspond to two different points of view the two test particles
approaching one another.

Note that, by demanding that the expression for $u/v$, as given by Eq. (9) and (10)
 must match at $r=2M$, one could have directly concluded that $u/v =\pm 1$
for $r=2M$.
 Of course, the fact that $u/v =1$ at the EH
also follows from the fact the EH corresponds to $t
=\infty$ (Eq. 23). However, the negative value $u/v =-1$ corresponds to $t
=-\infty$! This is again another anomaly associated with the assumption of
$M>0$. And finally all such exercises are actually unnecessary because
Eqs. (8) and (12) directly show that $u/v = \pm 1$ at the EH.

We have found that {\em for the Lemaitre
coordinate too},  $ds^2 =0$ at $r=2M$.
This implies that although the metric coefficients can be made to appear
 regular, the radial
geodesic of a {\em material particle becomes null} at the event
horizon of a finite mass BH in contravention of the basic
premises of GTR! And since, now, we can not blame the coordinate
system to be faulty for this occurrence, the only way we can
explain this result is that {\em the Event Horizon itself
corresponds to the physical singularity} or, in other words, the
mass of the Schwarzschild BHS $M\equiv 0$.

Finally, we offer here {\em the physical reason why the speed of
 free fall at the Schwarzschild radius must be equal to the
speed of light}. It is well known that the radial speed of a free
falling particle at $r=2M$, as measured by a {\em static}
Schwarzschild observer at $r=2M$ is $V_{sch}=1$. And this is
closely linked with the occurrence of the Schwarzschild
singularity. Let the radial speed at the Event Horizon in a
certain other coordinate system (such as Kruskal or Lemaitre or
any thing else) be $V$. And {\em we require that $\mid V\ <1$ at
$r=2M$ in order that EH can really occur at a finite value of}
$r$. And let {\bf locally} the relative velocity of the ``other
static observer'' be $V_{Sch-O}$. Now by principle of
equivalence, we can apply special theory of relativity locally,
and, thus at $r=2M$, we would have

\begin{equation}
V= {V_{sch} \pm V_{Sch-o} \over 1\pm V_{sch} V_{Sch-o}}
\end{equation}

And it is well known as well as trivial that that, in this case
we would always have $\mid V\ \equiv 1$ because $V_{sch}=1$ at
$r=2M$. And hence {\em there can not be any EH at a finite value
of $r$.}

 And then, the entire
conundrum of ``Schwarzschild singularity'', ``swapping of spatial
and temporal characters by $r$ and $t$ inside the event horizon
({\em when the angular part of all metrics suggest that $r$ has a
spacelike character even within the horizon}), ``White Holes''
and ``Other Universes'' get resolved. Similarly, the concepts of
``Worm Holes'' ceases to exist as well.

To conclude, irrespective of the observational consequences, we
have {\em directly} shown that, if GTR is correct, Schwarzschild
BHs must have $M\equiv 0$ in order that the radial geodesics of
material particles remain timelike at a finite value of $r$. And
physically this corresponds to the simple fact that the magnitude of
physical 3-velocity can not exceed the speed of light.

And we remind the reader that our result has been obtained from
three independent and different considerations, which so far,
nobody has been able to disprove\cite{3}. And we are thankful to Tereno
for making the first attempt to scientifically criticize our work.

\end{document}